\documentclass[preprint]{elsarticle}

\usepackage{amssymb}
\usepackage{amsmath}
\usepackage{colortbl}

\journal{}

\makeatletter
\def\ps@pprintTitle{%
  \let\@oddhead\@empty
  \let\@evenhead\@empty
  \def\@oddfoot{\reset@font\hfil\thepage\hfil}
  \let\@evenfoot\@oddfoot
}
\makeatother

\begin{document}

\begin{frontmatter}



\title{CAD Based Design Optimization of Four-bar Mechanisms: a coronaventilator case study}


\author[label1,label2]{Abdelmajid Ben Yahya \corref{cor1}}

\cortext[cor1]{Corresponding author\\
email adresses: Abdelmajid.Benyahya@uantwerpen.be (Abdelmajid Ben yahya),\\ Nick.VanOosterwyck@uantwerpen.be (Nick Van Oosterwyck),\\ Ferre.Knaepkens@uantwerpen.be (Ferre Knaepkens), Simon.Houwen@UGent.be (Simon Houwen), Stijn.Herregodts@UGent.be (Stijn Herregodts), Jan.Herregodts@UGent.be\\ (Jan Herregodts), Bart.Vanwalleghem@UGent.be (Bart Vanwalleghem),\\ Annie.Cuyt@uantwerpen.be (Annie Cuyt), Stijn.Derammelaere@uantwerpen.be (Stijn\\ Derammelaere)}

\affiliation[label1]{organization={Department of Electromechanics, CoSysLab, University of Antwerp},
            city={Antwerp},
            country={Belgium}}
\affiliation[label2]{organization={AnSyMo/CoSys, Flanders Make, the strategic research centre for the manufacturing industry},
            country={Belgium}
            }

\author[label1,label2]{Nick Van Oosterwyck}
\author[label3]{Ferre Knaepkens}
\author[label6,label7]{Simon Houwen}
\author[label5]{Stijn Herregodts}
\author[label5]{Jan Herregodts}
\author[label6,label7]{Bart Vanwalleghem}
\author[label3,label4]{Annie Cuyt}
\author[label1,label2]{Stijn Derammelaere}

\affiliation[label3]{organization={Department of Mathematics and Computer Science, University of Antwerp},
            city={Antwerp},
            country={Belgium}}
\affiliation[label4]{organization={College of Mathematics and Statistics, Shenzhen University Shenzhen},
            city={Guangdong 518060},
            country={China}}
\affiliation[label5]{organization={Department of Human Structure and Repair, Ghent University},
            city={Gent},
            country={Belgium}}
\affiliation[label6]{organization={Department of Electrical Energy, Metals, Mechanical Constructions and Systems, Ghent University Campus Kortrijk},
            city={Kortrijk},
            country={Belgium}}
\affiliation[label7]{organization={EEDT-MP, Flanders Make, the strategic research centre for the manufacturing industry},
            country={Belgium}}
            
\begin{abstract}

Design optimization of mechanisms is a promising research area as it results in more energy-efficient machines without compromising performance. However, machine builders do not actually use the design methods described in the literature as these algorithms require too much theoretical analysis. Moreover, the design synthesis approaches in the literature predominantly utilize heuristic optimizers leading to suboptimal local minima.

This research introduces a convenient optimization workflow allowing wide industrial adoption, while guaranteeing to reveal the global optimum. To guarantee that we find the global optimum, a mathematical expression of the constraints describing the feasible region of possible designs is of great importance. Therefore, kinematic analysis of the point-to-point (PTP) planar four-bar mechanism is discussed to obtain the static and dynamic constraints. Within the feasible region, objective value samples are generated through CAD multi-body software. These motion simulations determine the required torque to fulfill the movement for a certain combination of design parameters. Sparse interpolation techniques allow minimizing the required amount of samples and thus CAD simulations. Moreover, this interpolation of simulation results enables the representation of the objective in a mathematical description without in-depth analytical design analysis by the machine designer. Subsequently, the mathematical expression of the objective allows global optimizers to find a global optimal design within the feasible design space. In a case study of a coronaventilator mechanism with three design parameters (DP's), 1870 CAD motion simulations from which only 618 are used to build a model allowed to reduce the RMS torque of the mechanism by 67\%.

\end{abstract}




\end{frontmatter}


\section{Introduction} 

\label{sec:Introduction}

\begin{figure}[h]
    \centering
    \includegraphics[width=0.5\columnwidth]{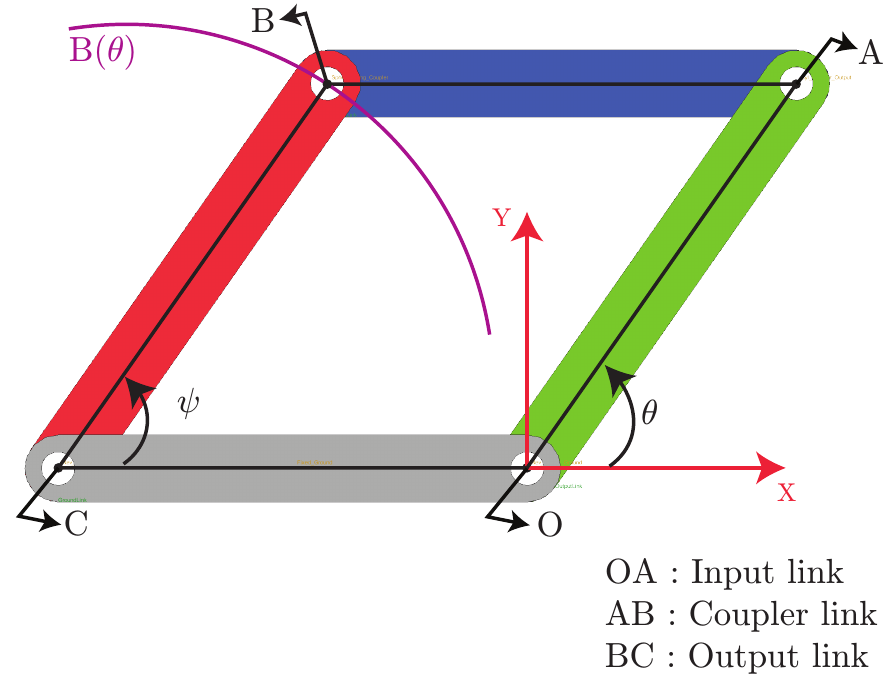}
    \caption{The considered design parameters $\vert OA \vert$, $\vert AB \vert$ and $\vert BC \vert$ of a four-bar, in the present paper.}
    \label{fig:fourbarDPs}
\end{figure}

The energy consumption of industrial machinery is a topic of primary importance due to environmental and economic considerations \cite{Dornfeld2013}. The 45\% share that electric motors have in the global electric consumption \cite{Waide2011} supports the statement that any energy-saving method should be investigated thoroughly. The methodology proposed within this paper is applicable for all planar four-bar mechanisms with an imposed movement of the end-effector and/or output link BC (see Figure \ref{fig:fourbarDPs}). The potential of this scope is indicated in \cite{Russell2005,Reis2014,Berselli2016}, stating that four-bar linkages are extensively used in practical engineering applications. Moreover, reciprocating PTP machinery is progressively common within the industry \cite{Berselli2016}. 

\begin{figure}[h]
    \centering
    \includegraphics[width=0.75\columnwidth]{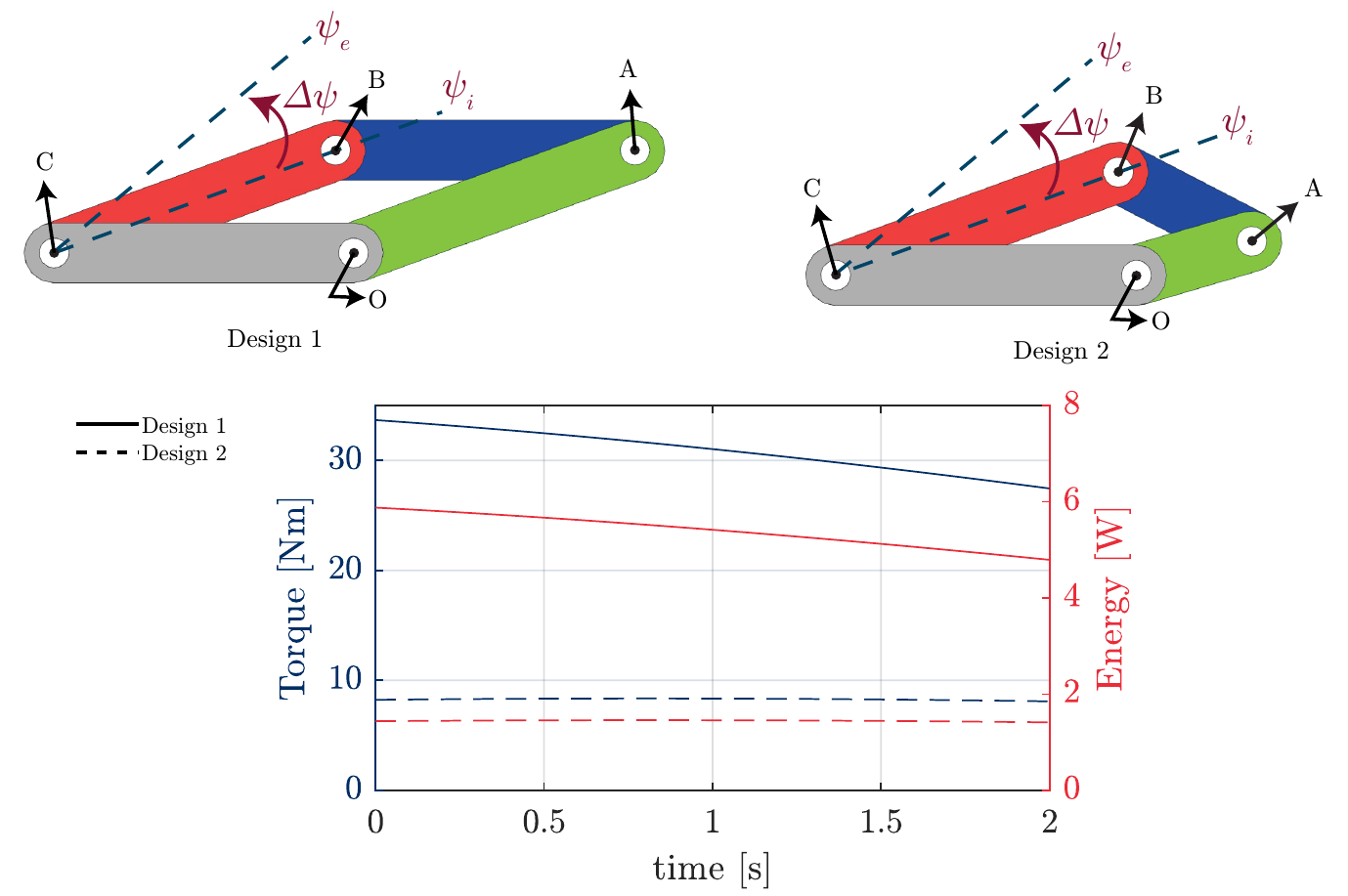}
    \caption{Defining certain lengths for the links of a four-bar influences the required torque to move the output link BC from $\psi_i$ to $\psi_e$ driven from point O / input link OA.}
    \label{fig:fourbar_optimconcept}
\end{figure}

The link lengths in a four-bar mechanism can differ while fulfilling the same task, being the PTP displacement of output link BC. Therefore, the geometry parameters depicted in Figure \ref{fig:fourbarDPs} can be considered as design parameters to enhance the mechanism. 
Design optimization of a Point-To-Point (PTP) mechanism is one specific approach to reduce the energy consumption of electric machinery, as indicated in Figure \ref{fig:fourbar_optimconcept}. Awareness about the influence of machine components geometry on energy consumption has recently attracted attention \cite{Carabin2017,Mashimo2015,Sheppard2019}. Mechanism models \cite{Oosterwyck2019,VanOosterwyck2020} replace the prototyping, allowing computational evaluation of multiple designs with limited costs.

A coronaventilator is used as a validation case within this study. This mechanism was constructed during the first wave of the covid-19 pandemic by a non-profit organization \cite{Herregodts2019}. Having continuous (24/7) access to electricity is not obvious within low- and middle-income countries. Thus, having a mechanism that consumes a minimum of electric energy enabling the usage of batteries is highly relevant. Therefore, the objective of this study is to find the optimal design (being lengths $\vert OA \vert$, $\vert AB \vert$ and $\vert BC \vert$) leading to a minimal $T_{RMS}$ for a reciprocal four-bar mechanism. The method introduced in this paper relies on CAD software to sample the objective function through motion simulations. These simulation tools allow broad industrial applicability. Furthermore, sparse interpolation technique is implemented to avoid an infeasible computational burden of numerous CAD simulations. Moreover, the objective function is only to be considered in the feasible design space of the four-bar mechanism.  Hence, constraints that limit the design space are highly relevant. State-of-the-art techniques generally use heuristic optimizers which cannot guarantee to find the global optimum \cite{Piazzi1998a}. However, the method described in this paper guarantees to reveal the global optimum.

\subsection{Related work}
\label{sec:Related work}
In the literature the minimisation of the driving torque is done by establishing dynamic equations of the system to predict the dynamics. However, this makes the method inconvenient for machine builders. Moreover, \cite{El-Kribi2013,Affi2007,Rayner2009} do not define the feasible search domain nor include it in searching for the optimum result. The constraints that define the feasible design space are important as defects, giving infeasible designs, \cite{Hernandez2021} frequently occur in the kinematic mechanism synthesis of a four-bar linkage. The optimization algorithms of \cite{El-Kribi2013,Affi2007,Gogate2012} assure that the objective function has converged towards a minimum, yet it is generally not guaranteed that the designed linkage will be feasible. Therefore, the necessary constraints should be added so that the optimal solution can fulfill the movement without inconveniences. Using a constrained-global optimization algorithm requires a deterministic mathematical description of the constraints to find the global optimum. To the authors’ knowledge, this has not been done yet in the literature \cite{Shen2015}.

Developing a four-bar mechanism that follows the desired output trajectory is a classic design problem that researchers extensively explore \cite{hrones1978,Jaiswal2017,Bai2015,Li2020,Li2016}. However, all methods above are not implementable in global optimizers as the algebraic expression (when provided) is only evaluated in discrete defined points $\begin{bmatrix} x_{B}(\theta) \\ y_{B}(\theta) \end{bmatrix}_{i \in \mathbb{N}}^*$ on the coupler curve \textbf{B}($\theta$) (shown in Figure \ref{fig:fourbarDPs}). Thus, these cannot deliver a deterministic mathematical description of the feasible design space, which is required.

\subsection{Method}
\label{sec:Method}

In general, it will be shown how CAD-based motion simulations combined with a sparse interpolation technique enable a global optimizer that guarantees revealing the global optimum and thereby outperform heuristic optimizers regarding energy savings.

Mechanical design of systems is mainly done in Computer-Aided Design (CAD) software. These CAD models include all required information (i.e., volume, mass, friction, damping, joints,...) to model the dynamics of a mechanism. This information is necessary to calculate the necessary torque of the mechanism through motion simulations. By driving the mechanism with the motion profile $\theta(t)$ at point O (Figure \ref{fig:fourbarDPs}), the location where the mechanism is driven in reality by a motor, the user can extract the necessary torque from the software (as in Figure \ref{fig:fourbar_optimconcept}) to fulfill the prescribed movement $\Delta\psi$ of the output link BC. Furthermore, within these motion simulations, the design parameters $\vert OA \vert$, $\vert AB \vert$ and $\vert BC \vert$ of the four-bar can be parameterized to simulate different designs. The objective value, to minimize by the optimizer is the RMS Torque (\(T_{RMS}\)) value, necessary to drive the mechanism fulfilling an imposed PTP motion ($\Delta\psi$). The literature states that minimizing the \(T_{RMS}\) corresponds with reducing the energy losses in the system \cite{Berselli2016}.

Hence, by calculating the RMS Torque based on CAD simulations as elucidated in Section \ref{sec:CAD motion simulations}, the objective value for a certain design (i.e., certain values for the three design parameters $\vert OA \vert$, $\vert AB \vert$ and $\vert BC \vert$) is obtained. The whole simulation process to obtain the objective value for different design parameter combinations ($\vert OA \vert$, $\vert AB \vert$ and $\vert BC \vert$) is automated. Constraints on the design parameter values are necessary to define an area containing feasible designs, as discussed in Section \ref{sec:Design parameter constraints}, from which designs are selected to simulate their corresponding objective value (\(T_{RMS}\)). As one design evaluation can take on average 1 minute and 25 seconds, computational simulation time becomes a burden. Therefore, wise selection of the simulated designs within the feasible design space is essential. The brute force method requires an inconceivable number of \(g^d\) motion simulations, with \textit{g} being the granularity of sampling and \textit{d} the number of design parameters. Even with state-of-the-art interpolation techniques, the construction of the objective function would require at least \((d+1).n^2.log^{2d-2}(n)\) samples \cite{Sauer2018}, with \textit{n} the total number of terms in the mathematical description of the objective function. In the case of the coronaventilator this would mean 782,933 samples are required. Therefore, the selection of samples is performed with certain rules in order to use an innovative multidimensional sparse interpolation approach \cite{Cuyt2018}. This novel interpolation technique, introduced in Section \ref{sec:Multidimensional sparse interpolation}, allows obtaining the objective function with a sparse sampling method within the feasible design space. This reduces the number of required samples to 618, with an additional 1252 validation samples. Limiting the number of samples to construct the objective function is a major enabler for a global optimizer. As the interpolation limits the number of CAD motion simulations to a bare minimum. In this case the number of necessary samples is reduced from 10,000,000 to 1870.

\section{CAD Motion Simulations}

\label{sec:CAD motion simulations}

\begin{figure}[h]
    \centering
    \includegraphics[width=0.5\columnwidth]{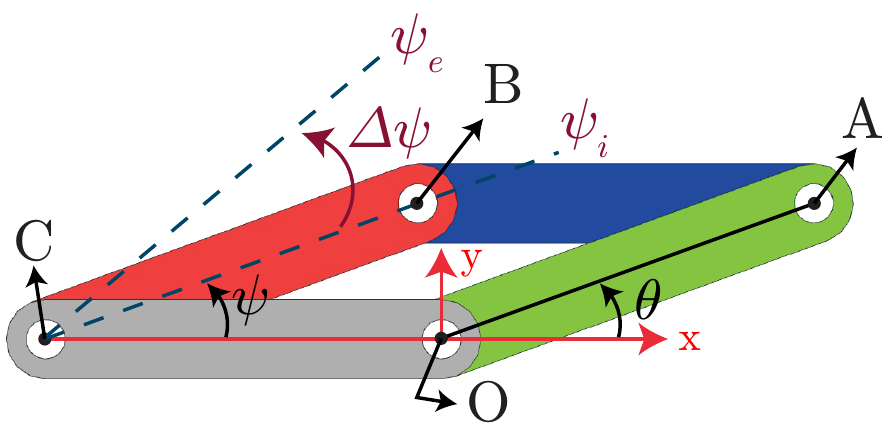}
    \caption{The output link BC requires a movement from $\psi_i$ to $\psi_e$, which is performed by moving $\theta$ over a design-specific angle.}
    \label{fig:outputmotion}
\end{figure}

In kinematic analysis, linkage dimensions $\vert OA \vert$, $\vert AB \vert$, $\vert BC \vert$ and $\vert OC \vert$ are known and the resulting output motion $\psi(t)$ (and its derivatives) can be calculated. On the other hand, dimensional synthesis is regarded as the inverse, in which for a specific output motion $\psi(t)$ the feasible dimensions of the linkages are obtained \cite{Lee2018a}. This paper is based on the dimensional synthesis of a planar four-bar function generation \cite{Lee2018}. As shown in Figure \ref{fig:outputmotion}, the movement $\Delta\psi$ of output linkage BC caused by $\theta(t)$ is described by a starting angle $\psi_i$, and end angle $\psi_e$. In this paper, the machine designer only defined an output motion $\psi(t)$ which results in a reciprocal movement between the positions $\psi_i$ and $\psi_e$.

\begin{figure}[h]
    \centering
    \includegraphics[width=1.0\columnwidth]{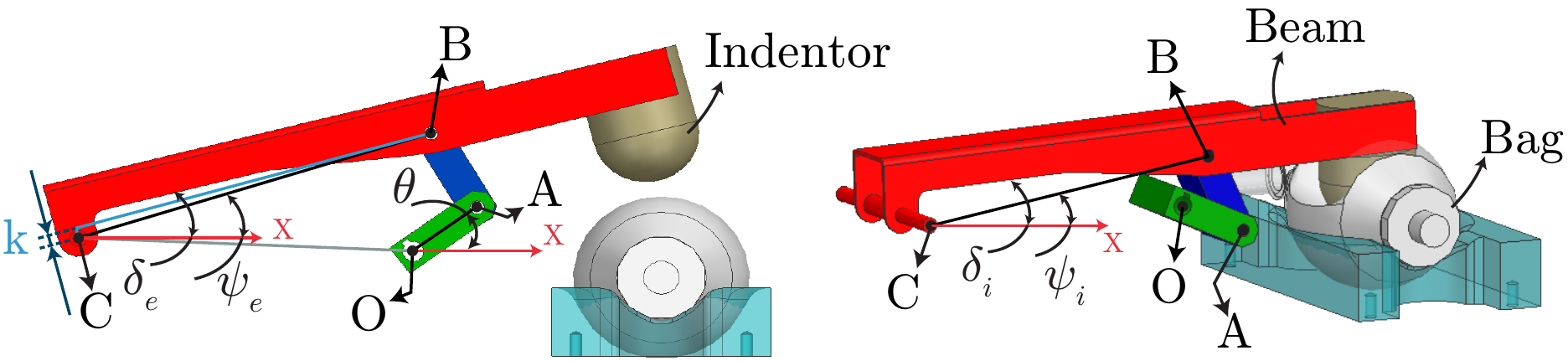}
    \caption{The used case within this research is a coronaventilator developed by Gear Up Medical vzw \cite{Herregodts2019}. It can be seen that the mechanism is from a four-bar type.}
    \label{fig:Coronaventilator}
\end{figure}

The validation case is clarified to make all the following more tangible. This mechanism, shown in Figure \ref{fig:Coronaventilator}, can ventilate a patient by pressing the indentor into the bag, which causes airflow towards the patient. Figure \ref{fig:Coronaventilator} presents the CAD model of the coronaventilator and illustrates that the red beam, connected with the indentor (i.e., the end-effector), moves by rotating input link OA around point O. This is the point where an electric motor drives the mechanism. The red beam has two predefined angles: an angle $\delta_e$ that holds the mechanism in a position where the indentor touches the bag and an angle $\delta_i$ that corresponds to a position in which the air is compressed out of the bag. Figure \ref{fig:Coronaventilator} clearly shows that the mechanism is a four-bar linkage on which the method proposed in this paper can be applied.

However, the output motion of the coronaventilator is described by the angle $\delta$ (linked to the red beam), while the four-bar has an angle $\psi$ that is linked to the output link BC. A relation between these angles stated as
\begin{equation}
    \psi=\mathrm{asin}\left(\frac{\sin\left(\delta\right)\,\left(\frac{k}{\mathrm{tan}\left(\delta\right)}+\sqrt{b^2-k^2}\right)}{b}\right)
    \label{eq:delta_psi}
\end{equation}
allows a conversion from $\delta$ to $\psi$. The parameter \textit{k} in Equation \eqref{eq:delta_psi} (Figure \ref{fig:Coronaventilator}) is a constant value that changes neither in the optimization nor during the four-bar mechanism's movement.

\begin{figure}[h]
    \centering
    \includegraphics[width=1.0\columnwidth]{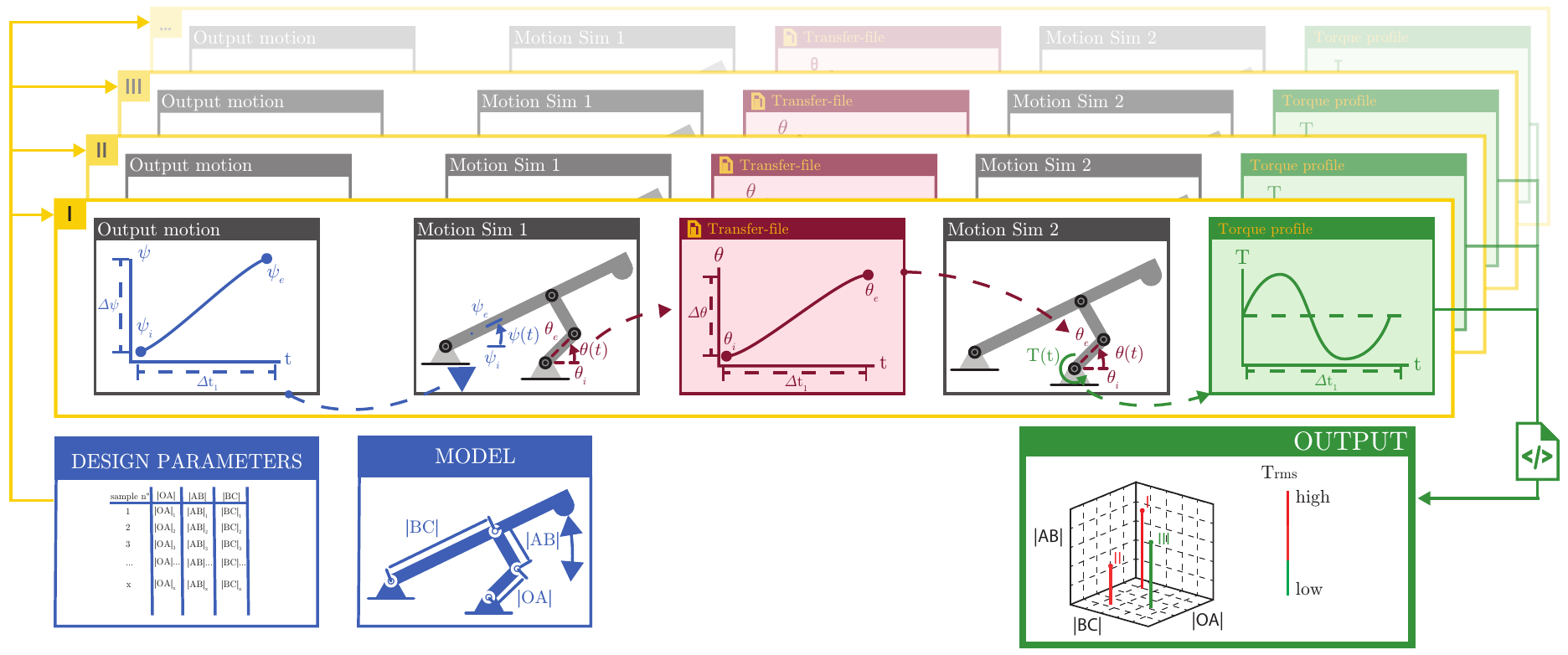}
    \caption{The approach performing the necessary driving torque calculation of the mechanism.}
    \label{fig:dynamic_analysis}
\end{figure}

A CAD motion simulation \cite{SiemensNX} can determine the necessary torque, to drive the mechanism at point O only if the required position profile $\theta(t)$, at that point O, is known. However, the user solely defines the required position profile of the end-effector, in this case $\delta(t)$. According to Equation \eqref{eq:delta_psi} we obtain $\psi(t)$. The conversion of $\psi(t)$ to $\delta(t)$ depends on the values of the design parameters $\vert OA \vert$, $\vert AB \vert$ and $\vert BC \vert$. Therefore, each selected design is analysed by two motion simulations, as indicated in Figure \ref{fig:dynamic_analysis}. 
If the design is combined with the required output motion $\psi(t)$, the first kinematic motion simulation can extract the required motor position displacement $\theta(t)$. Subsequently, the motor motion profile $\theta(t)$ is used in the second motion simulation, from which the required driving torque is determined. This process with kinematic simulation and subsequent torque calculation extracts the objective value for predefined designs.

\section{Design Parameter Constraints}

\label{sec:Design parameter constraints}

\subsection{Static constraints of a four-bar}
\label{sec:Static_constraint}
\begin{figure}[h]
    \centering
    \includegraphics[width=0.5 \columnwidth]{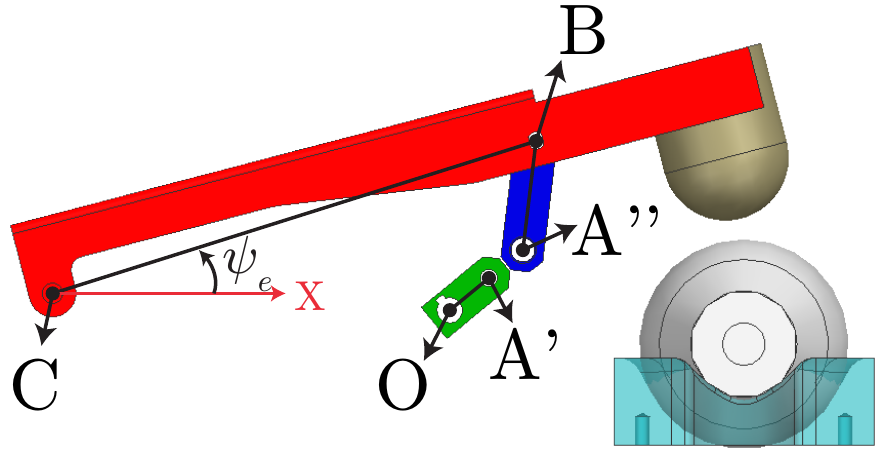}
    \caption{The combination of design parameters $\vert OA \vert$, $\vert AB \vert$ and $\vert BC \vert$ serve an infeasible design that cannot be assembled in $\psi_{e}$.}
    \label{fig:Infeasible_design}
\end{figure}

\begin{figure}[h]
    \centering
    \includegraphics[width=0.5 \columnwidth]{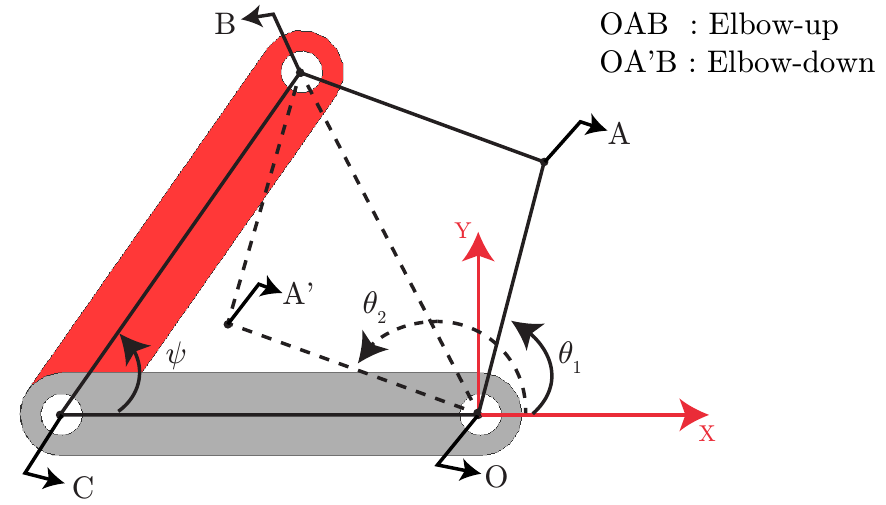}
    \caption{The elbow-up OAB and elbow-down OA'B are two possible constructions in which the four-bar linkage can be assembled for one $\psi$ angle of the output link BC.}
    \label{fig:elbowconfig}
\end{figure}

The combinations of design parameters $\vert OA \vert$, $\vert AB \vert$ and $\vert BC \vert$, to consider in the workflow above, are chosen so that the designs are located within the feasible design space of the four-bar mechanism. To determine this region of feasible designs, the first step is looking for static constraints. This implies that only the designs which can be assembled for the maximal and minimal angle of the output link BC ($\psi_i$ and $\psi_e$) can be part of the feasible design space. An example of a design that cannot be assembled in $\psi_{e}$ due to the chosen values for DP's $\vert OA \vert$, $\vert AB \vert$ and $\vert BC \vert$ is illustrated in Figure \ref{fig:Infeasible_design}. This shows that the input link OA' cannot be connected with the coupler link A''B. 
\begin{figure}[h]
    \centering
    \includegraphics[width=1 \columnwidth]{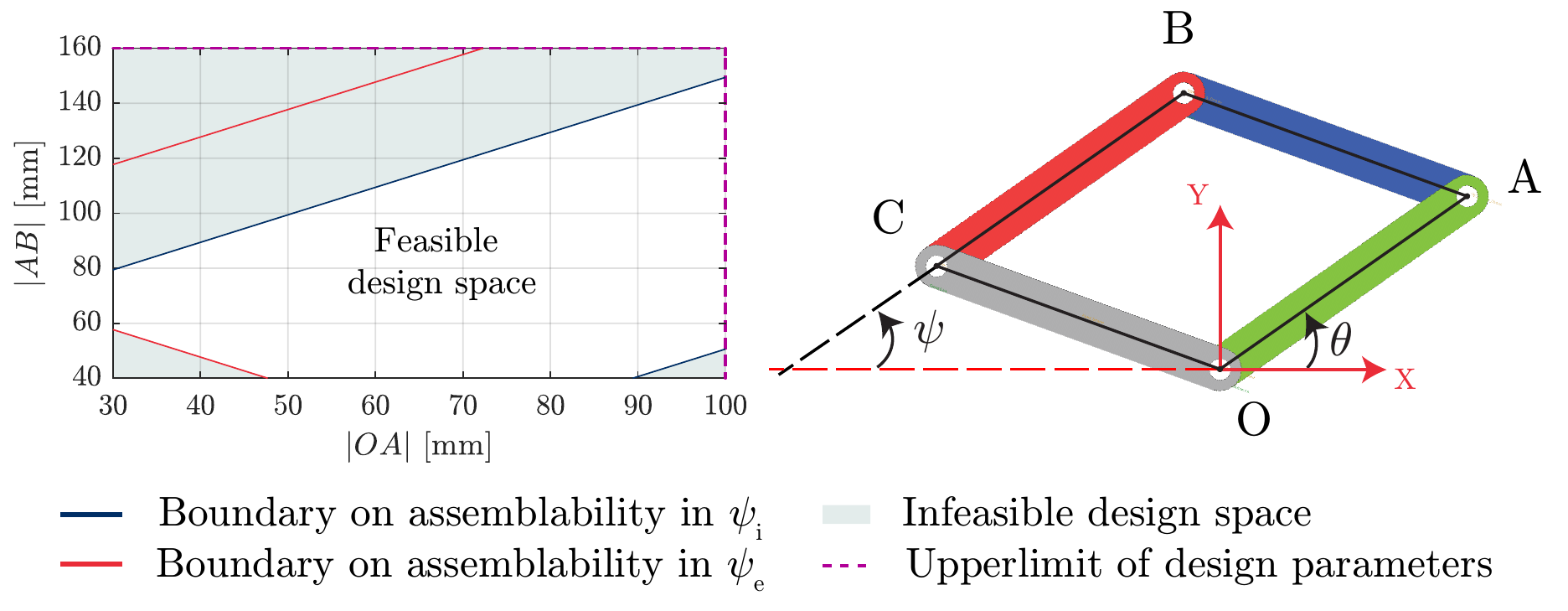}
    \caption{The static constraints (left), shown for 2 DP's, limit the design space to the area of designs that reach the output link's (BC) start- and end position ($\psi_i$ and $\psi_e$).}
    \label{fig:Static_constraint}
\end{figure}
This assemblability define the first boundaries on the design space that is illustrated in Figure \ref{fig:Static_constraint} (left) (only in 2D for illustrative purposes). These boundaries are obtained through a position analysis of the four-bar mechanism for both begin $\psi_{i}$- and end-position $\psi_{e}$. For analysis of the ventilator, the origin of the fixed frame is placed in joint O (the driver joint). Let $\theta$ be the angle of the input link OA measured relative to the x-axis and $\psi$ the angle of the output link BC relative to the x-axis, Figure \ref{fig:Static_constraint} (right). A relation between the input angle $\theta$ and output angle $\psi$ is obtained based on the length of the coupler link $\vert AB \vert$, which stays fixed during the mechanism's movement \cite{MCCarthy2010}. Therefore, the analysis can start with:
\begin{equation}
\Bigg ( \begin{bmatrix} x_{B}(\psi) \\ y_{B}(\psi) \end{bmatrix}-\begin{bmatrix} x_{A}(\theta) \\ y_{A}(\theta) \end{bmatrix} \Bigg ).\Bigg ( \begin{bmatrix} x_{B}(\psi) \\ y_{B}(\psi) \end{bmatrix}-\begin{bmatrix} x_{A}(\theta) \\ y_{A}(\theta) \end{bmatrix} \Bigg )=|AB|^2
    \label{eq:Fixed_distance}
\end{equation}
where
\begin{equation}
\begin{split}
\begin{matrix}
    &x_{A}\left(\theta \right)=\vert OA \vert\, \cos\left(\theta \right) \\
    &y_{A}\left(\theta \right)=\vert OA \vert\,\sin\left(\theta \right)\\
\end{matrix}
\quad
\begin{matrix}  
    &x_{B}\left(\psi \right)=x_{C}+\vert BC \vert\,\cos\left(\psi \right)\\
    &y_{B}\left(\psi \right)=y_{C}+\vert BC \vert\,\sin\left(\psi \right).
\end{matrix}
\end{split}
\label{eq:Coordinates}
\end{equation}
By substitution of \eqref{eq:Coordinates} in \eqref{eq:Fixed_distance}, the dependency of the input angle $\theta$ based on the output angle $\psi$ is noted as
\begin{equation}
    \theta_{1,2}\left(\psi \right)= \mathrm{atan2}\left(V\left(\psi \right),U\left(\psi \right)\right) \pm \mathrm{arccos}\left(\frac{W\left(\psi \right)}{\sqrt{U^2\left(\psi \right)+V^2\left(\psi \right)}}\right) +\pi
    \label{eq:theta_for_psi}
\end{equation}
where
\begin{equation}
\begin{split}
    U\left(\psi \right)=&-2\,x_{C}\,\vert OA \vert-2\,\vert OA \vert\,\vert BC \vert\,\cos\left(\psi \right) \\
    V\left(\psi \right)=&-2\,y_{C}\,\vert OA \vert-2\,\vert OA \vert\,\vert BC \vert\,\sin\left(\psi \right)\\
    W\left(\psi \right)=&{x_{C}}^2+{y_{C}}^2+\vert OA \vert^2+\vert BC \vert^2-\vert AB \vert^2+2\,\cos\left(\psi \right)\,x_{C}\,\vert BC \vert\\
                        &+2\,\sin\left(\psi \right)\,y_{C}\,\vert BC \vert.
\end{split}
\label{eq:parts_of_theta_for_psi}
\end{equation}
Equation \eqref{eq:theta_for_psi} allows the derivation of the input angle $\theta$ from the output angle $\psi$. The latter is the imposed output motion defined by the $\Delta\psi$ range. However, the mechanism can be assembled in two ways for a single output angle $\psi$, resulting in two possible solutions for $(\theta)$ in Equation \eqref{eq:theta_for_psi}. This is a consequence of having the possibility to construct the four-bar, with a certain angle $\psi$, with output link BC on both sides of the diagonal OB, as shown in Figure \ref{fig:elbowconfig}. Both constructions, called the elbow-up OABC and elbow-down OA'BC, provide feasible solutions. The method proposed in the present paper applies to both configurations, yet it is chosen to focus on the elbow-up OABC, as it is the most efficient one according to \cite{Srivatsan2013}. To obtain the corresponding $\theta_{1}$ which is smaller than $\theta_{2}$, the second term is subtracted from the first term in Equation \eqref{eq:theta_for_psi}.

Regardless of the elbow configuration, feasibility constraints due to the solvability of Equation \eqref{eq:theta_for_psi} arise. A solution can be found if the argument of the $\arccos$ is in the range [-1,1]. Thus, a solution solely exist when the inequality constraint:
\begin{equation}
    U^2\left(\psi \right)+V^2\left(\psi \right)-W^2\left(\psi \right) \geq 0
    \label{eq:general_Static_constraint}
\end{equation}
is satisfied. In this way, an inequality constraint for the two output angles $\psi$ that bring point B farthest and closest to O must be established. Therefore, Equation \eqref{eq:general_Static_constraint} is evaluated for the maximal and minimal angle $\psi_{e}$ and $\psi_{i}$. This evaluates the assemblability in the positions $\psi_i$ and $\psi_e$.
\begin{equation}
    U^2\left(\psi \right)+V^2\left(\psi \right)-W^2\left(\psi \right)\bigg|_{\psi=\psi_{i},\psi_{e}} \geq 0
    \label{eq:Static_constraint_psi}
\end{equation}
By fulfilling Equation \eqref{eq:Static_constraint_psi}, one can say that the designed mechanism is assemblable over its movement. This design lies than within the area formed by the blue lines, which means that the mechanism is assemblable in $\psi_{i}$, and inside the area formed by red lines as it is assemblable in $\psi_{e}$ (see Figure \ref{fig:Static_constraint}).

\subsection{Dynamic constraints of a four-bar}
\label{sec:Dynamic_constraint}
\begin{figure}[h]
    \centering
    \includegraphics[width=1.0\columnwidth]{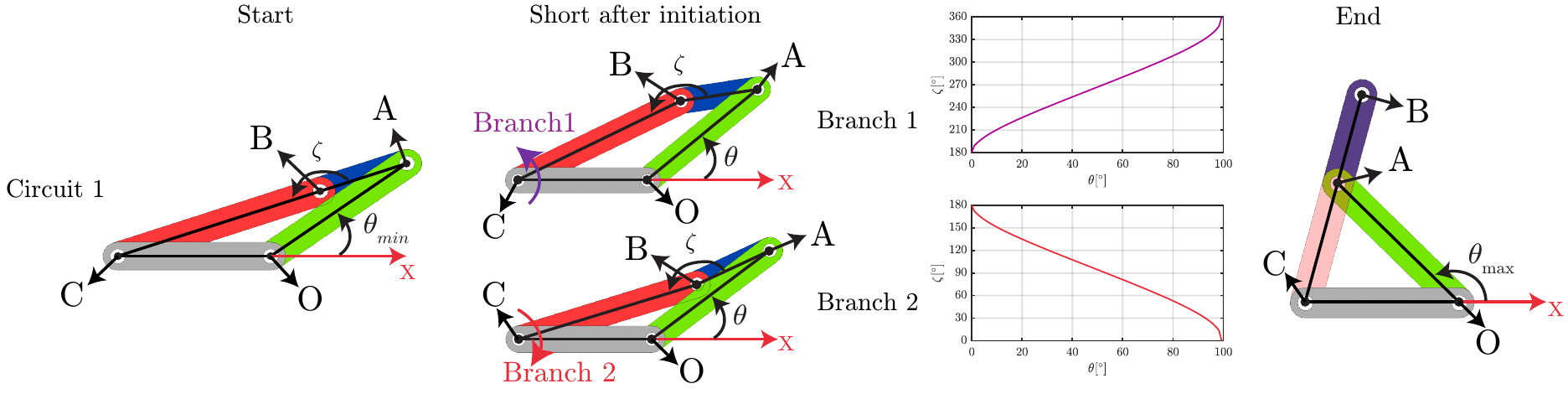}
    \caption{An example of a four-bar mechanism that has two connecting branches on the first circuit. It is shown that moving the mechanism from $\theta_{min}$ until $\theta_{max}$ corresponds with a movement of the transmission angle on branch 1 from 180° to 360° while on branch 2 from 180° to 0°.}
    \label{fig:fourbar_onecircuit}
\end{figure}

The aforementioned static constraints in chapter \ref{sec:Static_constraint} are not sufficient to exclude all infeasible designs. To ensure that the desired movement $\psi(t)$ of the output linkage BC is feasible, all defects during the movement should also be excluded. The three types of defects that can occur during the motion of a four-bar linkage are \textbf{branch}, \textbf{order} and \textbf{circuit} defects. The broad review in \cite{Balli2002a} reveals that research about branch, order and circuit defect avoidance is of great significance in the field of linkage synthesizes. With a \textbf{branch} defect, the mechanism cannot perform the desired movement continuously. More specifically, four-bar linkages can move in two different ways. In Figure \ref{fig:fourbar_onecircuit}, the input link OA moves between its extreme positions ($\theta_{min}$ until $\theta_{max}$), resulting in a change of the transmission angle $\zeta$ between 0 and $\pi$. The extreme input angle positions $\theta_{min}$ and $\theta_{max}$ corresponds with an angle $\zeta$ equal to respectively $\pi$ and 0. The movement is conducted by initiating the motion of the output link BC clockwise or counter-clockwise around C. The movement in each initial direction around C (clockwise or counter-clockwise) represents separate a branch. If the mechanism has to change branch while moving, a branch defect occurs for this linkage system design \cite{Singh2017}. When a branch defect occurs, one can observe that the mechanism reaches the $\theta_{min}$ or $\theta_{max}$ position more than once during the movement. This results in a transmission angle $\zeta$ moving through 0 or $\pi$. Hence, when the mechanism moves through the positions $\zeta$ equal to 0 or $\pi$, a change in the direction of $\theta$ occurs.

\begin{figure}[h]
    \centering
    \includegraphics[width=1.0\columnwidth]{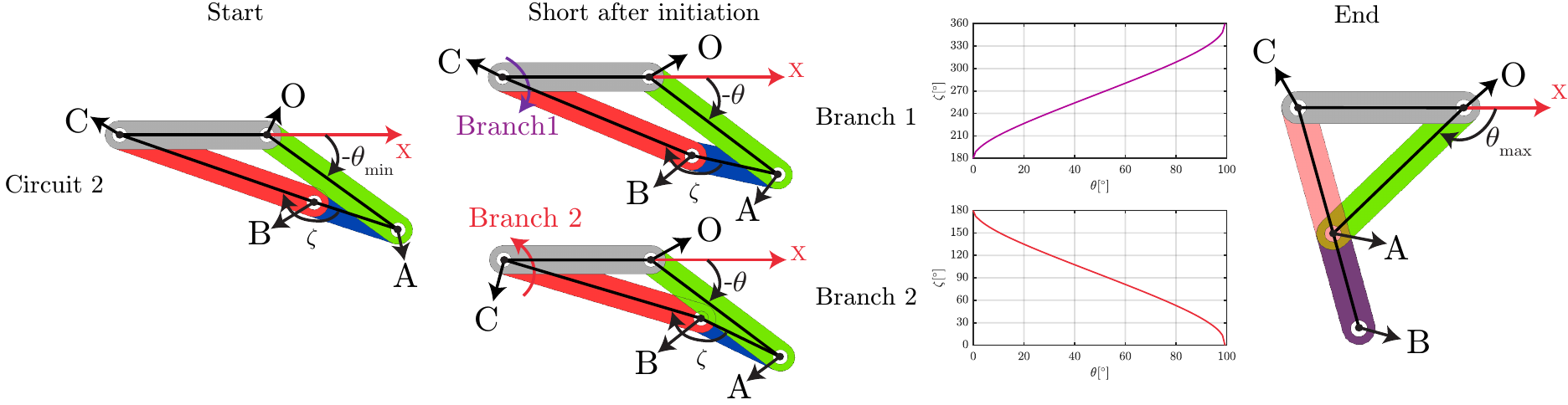}
    \caption{The second circuit of this specific four-bar design indicate that another circuit entails a complete different range. Nonetheless, the circuit is also constructed by two connected branches, with the same transition conditions for $\zeta$.}
    \label{fig:fourbar_circuittwo}
\end{figure}

\begin{figure}[h]
    \centering
    \includegraphics[width=0.8\columnwidth]{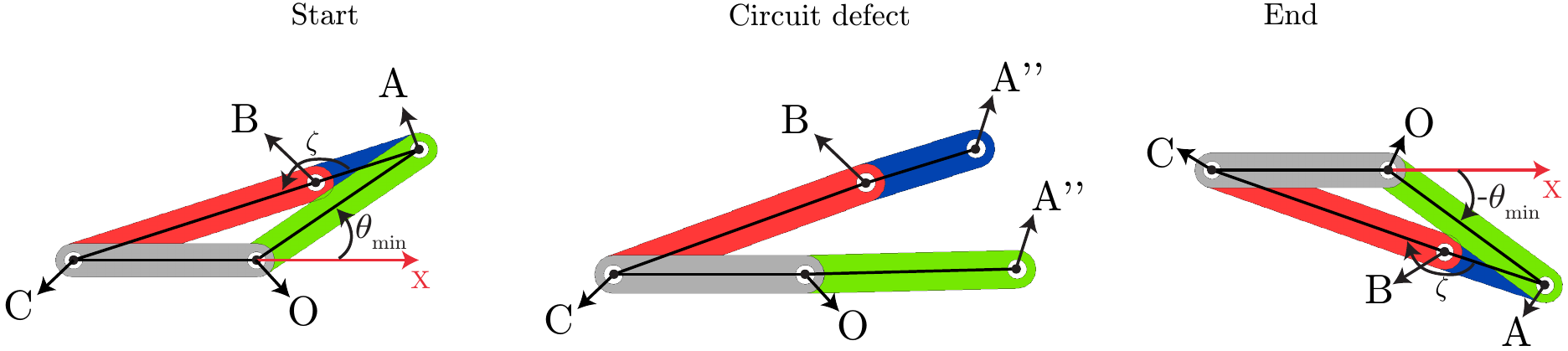}
    \caption{A designed linkage system that moves from one circuit to another must be disassembled, which is a circuit defect.}
    \label{fig:fourbar_circuitdefect}
\end{figure}

\textbf{Order} defects appear if certain points $\begin{bmatrix} x_{B}(\theta) \\ y_{B}(\theta) \end{bmatrix}_{i \in \mathbb{N}}^*$ are not reached in a certain sequence or order \cite{Gogate2012}. Order defects are impossible in this study as a reciprocal mechanism is considered, which moves continuously (with a fixed motion profile $\pi(t)$) between the maximal and minimal angle $\psi_{e}$ and $\psi_{i}$. 
Figure \ref{fig:fourbar_onecircuit} indicates that a \textbf{circuit} can exist out of two connected branches. Moreover, this design reveals that a mechanism can have an other circuit in which the mechanism reaches whole other positions $\begin{bmatrix} x_{B}(\theta) \\ y_{B}(\theta) \end{bmatrix}_{i \in \mathbb{N}}^*$, as shown in Figure \ref{fig:fourbar_circuittwo}. The maximum circuits a four-bar mechanism can have are limited to two.  The mechanism can move in each circuit separately without the necessity of disconnecting any joints \cite{chase1993}. A \textbf{circuit} defect arises when the linkage mechanism must be disassembled and placed in the other circuit, shown in Figure \ref{fig:fourbar_circuitdefect}, to complete the motion. To obtain a circuit defect, $\theta$ should become bigger or smaller than $\theta_{max}$ (with $\zeta$=0) or $\theta_{min}$ (with $\zeta$=$\pi$) respectively, to fulfill the desired movement of the output link BC ($\psi(t)$). A circuit defect has the same influence on $\theta$ as during a branch defect. In this paper, PTP movements with only a desired start- and endpoint are considered. The movement takes place through the actuation of one joint, point O. Therefore, the movement should stay within a single branch of a single circuit \cite{Feki2013} (Figure \ref{fig:fourbar_onecircuit} or \ref{fig:fourbar_circuittwo}). 

\begin{figure}[h]
    \centering
    \includegraphics[width=0.8\columnwidth]{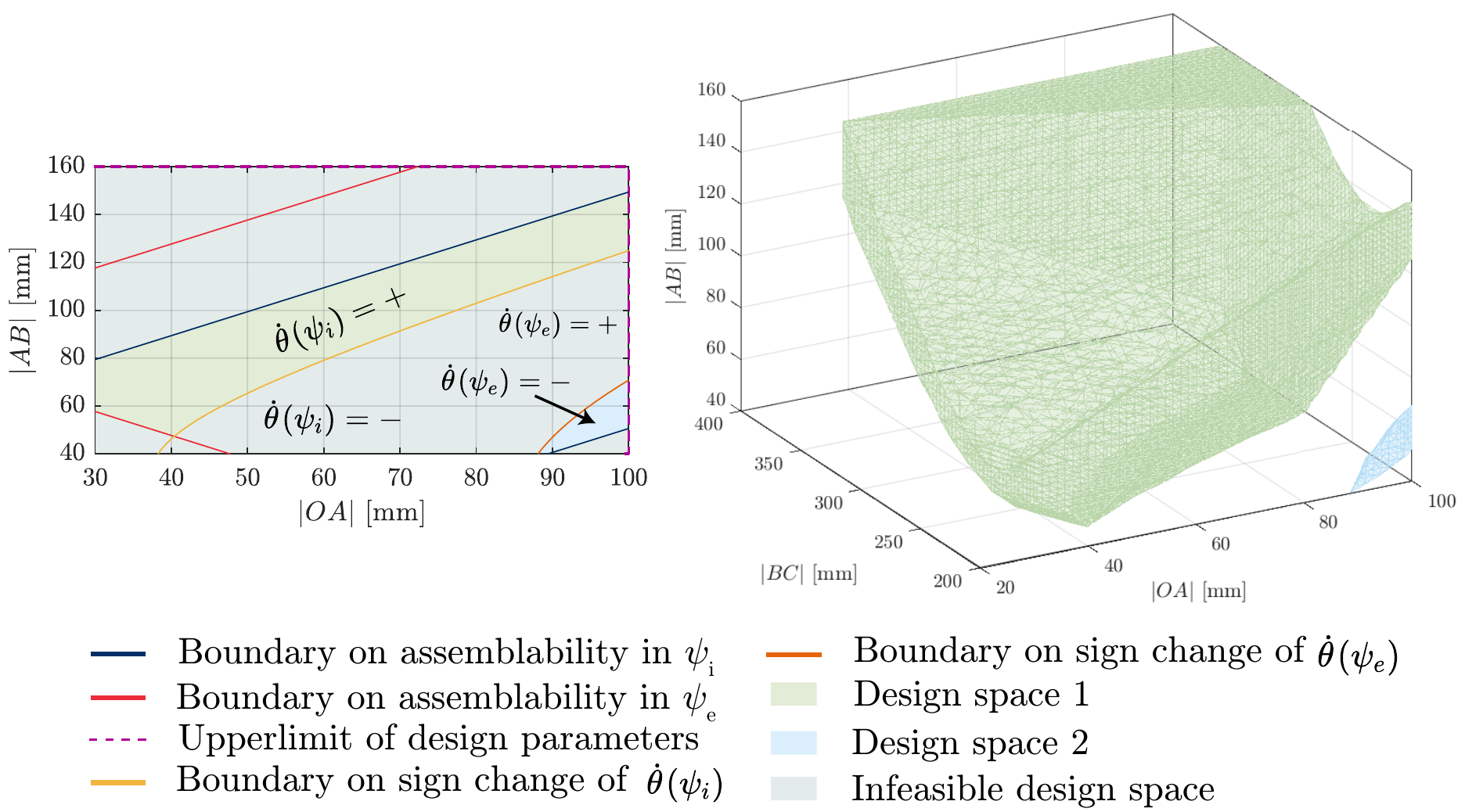}
    \caption{All constraints are shown for 2 DP's (left) and 3 DP's (right). The objects on the right are the feasible design spaces where the designs can perform the imposed reciprocal movement.}
    \label{fig:Constraints}
\end{figure}

To eliminate the possible circuit and branch defects, dynamic constraints are created based on the consequence of a defect that changes the direction of the input angle $\theta$. The calculations of the motor angle are always chosen for the elbow-up OABC. However, by altering the circuit, the configuration becomes an elbow-down in which $\theta$ moves in the other direction. Therefore, one can exclude branch and circuit defects by guaranteeing monotonicity in the motor position profile $\theta(t)$. The dynamic constraint 
\begin{equation}
sign\left(\dot{\theta}\left(\psi_{i} \right) \right)=sign\left(\dot{\theta}\left(\psi_{e} \right) \right)
    \label{eq:Dynamic_constraint_psi}
\end{equation}
will check if the first derivative of $\theta$, in the start- and end-position $\psi_{i}$ and $\psi_{e}$, does not alter its sign. Equation \eqref{eq:Dynamic_constraint_psi} is only applicable if the mechanism deals with an odd number of branch and or circuit defects while moving, as only then a change of sign is detected. Nonetheless, the method is still applicable when an even number of defects occur, because a defect results in very high required driving torques for each sign change. The interpolation in chapter \ref{sec:Multidimensional sparse interpolation} neglects these disproportional objective values. In that way, an even number of sign changes caused by an even number of defects will not affect the optimization. So, all the constraints together indicate the feasible design spaces, as shown in Figure \ref{fig:Constraints}. Within the feasible design spaces motion simulations, for certain samples, are performed to determine the objective value.

\section{Multidimensional Sparse Interpolation}

\label{sec:Multidimensional sparse interpolation}

Determining the objective values, using CAD multi-body simulations, for the three-dimensional design problem, in a brute force way would lead to a tremendous computational burden requiring 10,000,000 samples. Therefore, we rely on a sparse data fitting method to determine a mathematical model for the objective function $T_{RMS}(|OA|, |AB|, |BC|)$. While several interpolation methods are characterized by a trade-off between model accuracy and computing cost, sparse interpolation does not involve such compromise. The method introduced here uses a divide-and-conquer approach \cite{Briani2020, Cuyt2018} by splitting up the involved numerical linear algebra problems into smaller and hence better-conditioned independent sub-problems.

\begin{figure}[h]
    \centering
    \includegraphics[width=0.49\columnwidth]{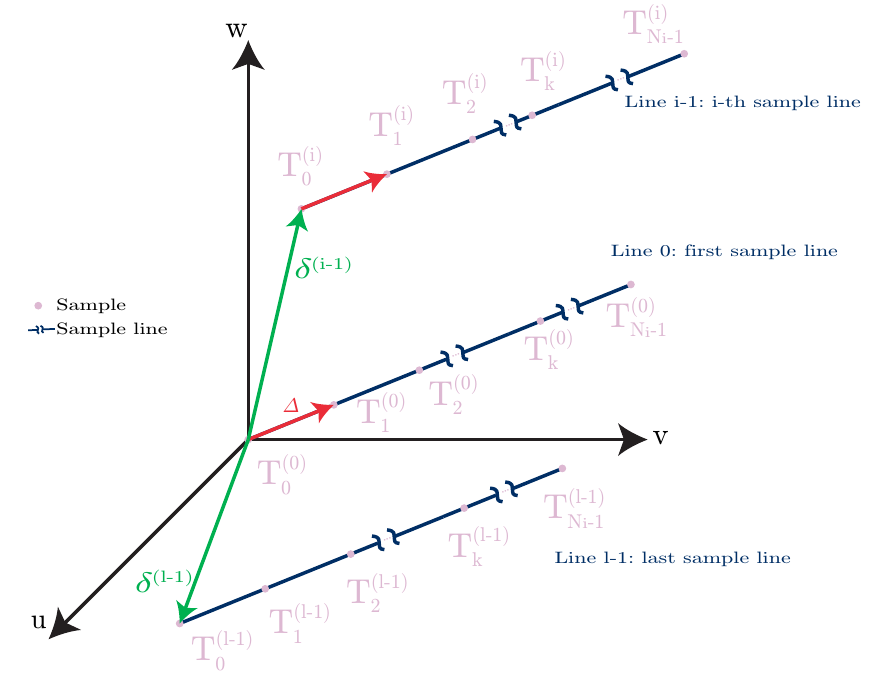}
    \includegraphics[width=0.49\columnwidth]{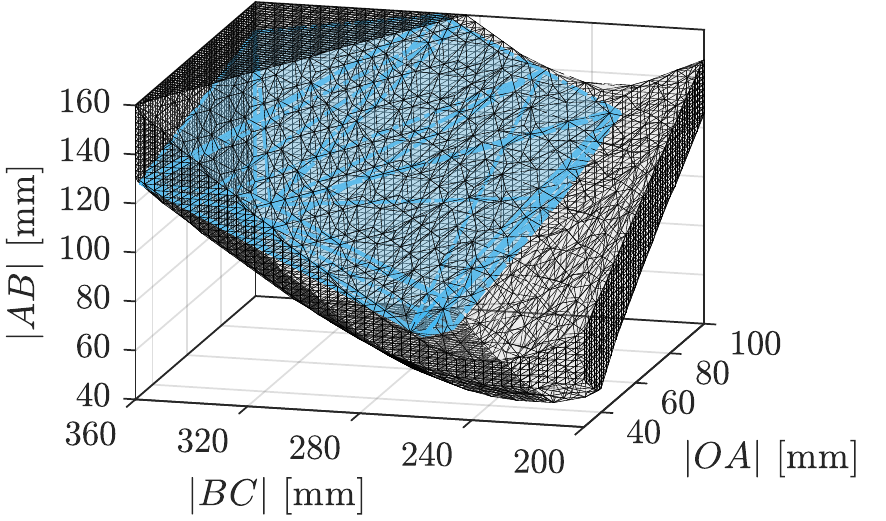}
    \caption{At the left, all the samples located at points $k\Delta + \delta^{(i)}$. The convex hull, in blue, of the $l$ lines covering the larger part of the design space, at the right.}
    \label{fig:samples and convexhull}
\end{figure}

For this method, the objective value, within the feasible design space as defined in \eqref{eq:Static_constraint_psi} and \eqref{eq:Dynamic_constraint_psi}, is determined on $l$ distinct lines in 3D space that are all parallel to a chosen vector $\Delta=(\Delta_u, \Delta_v, \Delta_w)$. We let $\delta^{(i)}, i=0, \ldots, l-1$ indicate the 3D vector that the $i$-th parallel line is shifted
over with respect to the line through the origin spanned by $\Delta$ for which we take $\delta^{(0)}=0$. Then the equidistant samples on these parallel lines, as depicted in Figure \ref{fig:samples and convexhull} left, are denoted by:
$$T_k^{(i)} := T_{RMS}(k\Delta+\delta^{(i)}), \qquad i=0, \ldots, l-1, \qquad
k=0, \ldots, N_i-1.$$
Let us compactly denote the tuple of design variables $(|OA|, |BC|, |AB|)$ by 
$$U= (u, v, w) := (|OA|, |BC|, |AB|)$$
and let $\langle \cdot, \cdot \rangle$ denote the standard inner product in 3D space. On each $i$-th parallel line the samples $T_k^{(i)}$ can be modeled by the sparse interpolant 
\begin{equation}
    T_{RMS,i}( U) = \sum_{j=1}^{n_i} \beta_j^{(i)} \exp \left( \langle \phi_j^{(i)},
    U \rangle \right)
    \label{eq:Sparse_interpolant}
\end{equation}
satisfying
$$T_k^{(i)} = \sum_{j=1}^{n_i} \beta_j^{(i)} \exp \left( k \langle \phi_j^{(i)},  
\Delta \rangle \right), \qquad i=0, \ldots, l-1, \qquad k=0, \ldots, N_i-1.$$
Note that the effect or influence of $\delta^{(i)}$ is absorbed into the coefficients $\beta_j^{(i)}$ in $T_{RMS,i}(U)$ which models the behaviour of $T_{RMS}$ on the $i$-th line.

The model for $T_{RMS,i}( U)$ can be computed using any of the existing 1D exponential
fitting methods, such as \cite{Briani2020,Roy_1989, Schmidt_1986,Steedly1994}. The number of terms $n_i$ in the sparse model can differ on each $i$-th line. The $l$ individual models are only valid on their respective line spanned by $\Delta$ and shifted over $\delta^{(i)}$. Now we need to blend these individual sparse models into an overall sparse model, valid in the convex hull (blue area in Figure \ref{fig:samples and convexhull} right) of the $l$ lines, which should cover the larger part of the region of interest. This requisite or demand actually dictates the more proper choices for $\Delta$ and the $\delta^{(i)}, i=1, \ldots, l-1$.

\begin{figure}[h]
    \centering
    \includegraphics[width=1.0\columnwidth]{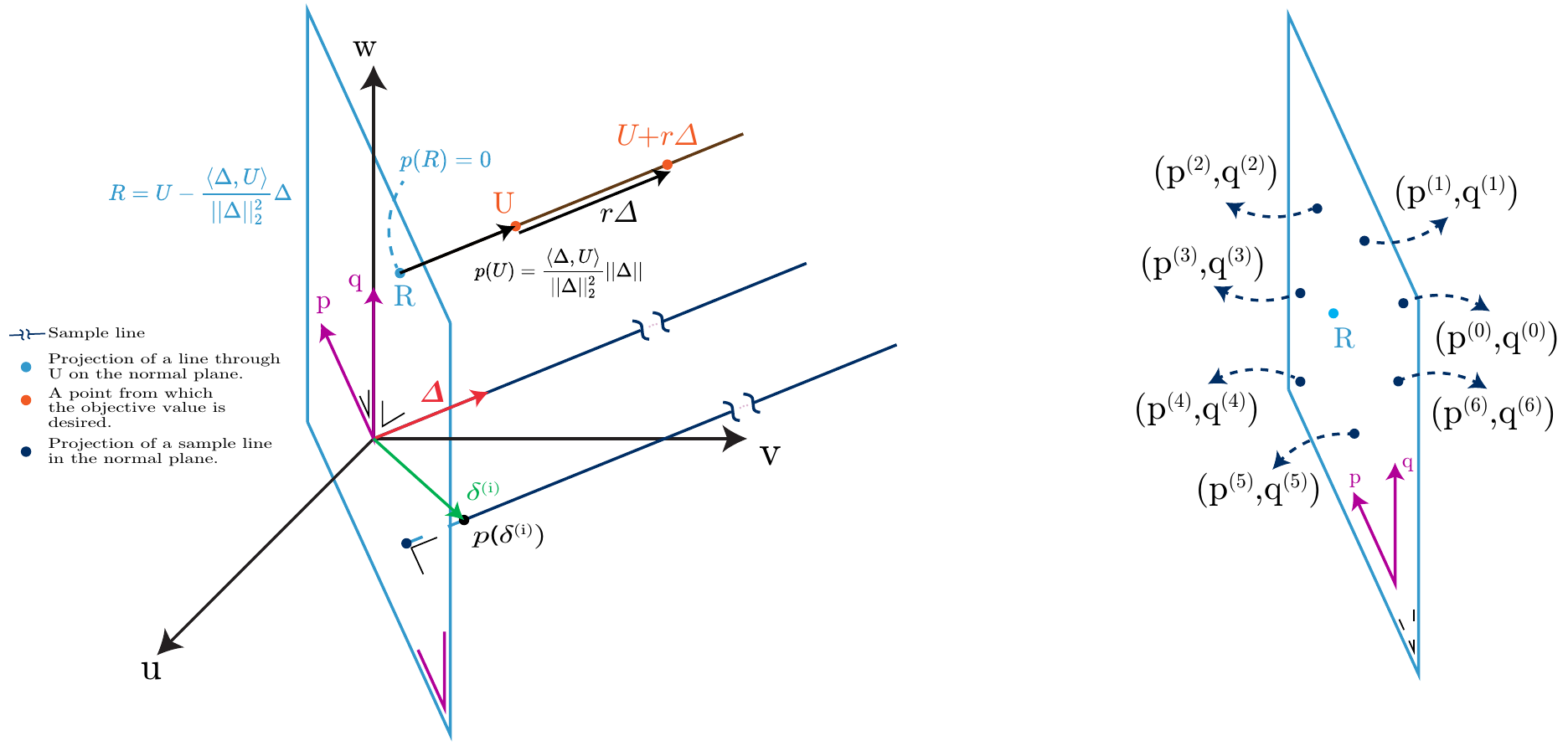}
    \caption{The intersection point $R$ of the line $U+r\Delta$, parallel with $\Delta$ and through $U$, at the left. All $l=7$ lines' intersection points $(p^{(i)}, q^{(i)})$ with the normal plane at the right.}
    \label{fig:intersection point}
\end{figure}

In what follows, we consider every design parameter combination $U$ in 3D space to lie on some line parallel with the one spanned by $\Delta$, also if $U$ is not an
interpolation point. All the points on such a line take the form
\begin{equation} 
    U + r\Delta, \qquad r \in \mathbb{R}.
\label{eq:line through point}
\end{equation}
The normal plane through the origin and orthogonal to $\Delta$ is given by the equation
$$\Delta_u u + \Delta_v v + \Delta_w w = 0,$$
or more compactly 
$$\langle \Delta, U \rangle = 0.$$
The intersection point $R$ of the normal plane with \eqref{eq:line through point} is thus given by
\begin{equation*}
    \langle \Delta, R \rangle = 0, \qquad R=U+r\Delta,
\end{equation*}
or more explicitly,
\begin{equation}
    R = U - \frac{\langle \Delta, U \rangle}{||\Delta||^2} \Delta.
    \label{eq:Intersection_point}
\end{equation}
Hence, the distance of $U$ to this intersection point $R$, expressed as a multiple of $||\Delta||$, equals 
$$p(u,v,w) = \frac{\langle \Delta, U \rangle}{||\Delta||^2}, \qquad U=(u, v, w)$$
and the points on the line given by \eqref{eq:line through point} can be re-expressed as
\begin{equation}
    R + p(U) \Delta
\end{equation}
On each line through a point $U$ parallel with $\Delta$, the intersection point $R$ with the normal plane, shown in Figure \ref{fig:intersection point}, is where $p(R)=0$ on the line. 

We therefore propose a blended 3D model of the following form, to represent the objective value $T_{RMS}$ overall, 
\begin{equation}
    T_{RMS}(u,v,w) \approx
    \sum_{i=0}^{l-1} \sum_{j=1}^{n_i} b_j^{(i)}(u,v,w) \exp \left( 
    p(u,v,w) \langle \phi_j^{(i)}, \Delta \rangle \right),
    \label{eq:3D_model_2}
\end{equation}
where the parameters $\phi_j^{(i)}$ and the value of $p(u, v, w)$ are already determined and where furthermore
the overall model continues to interpolate the values $T_k^{(i)}$ in the sample points $k\Delta+\delta^{(i)}$.
The blended model \eqref{eq:3D_model_2} coincides with the 1D models \eqref{eq:Sparse_interpolant} on each parallel line, and in between the lines the exponential terms 
fade in and out. Since 
$$p(k\Delta+\delta^{(i)}) = k + p(\delta^{(i)})$$
this means
\begin{equation}
    T_{RMS}(k\Delta+\delta^{(i)}) = \sum_{j=1}^{n_i} b_j^{(i)}(k\Delta+\delta^{(i)}) \exp \left( p(\delta^{(i)}) \langle \phi_j^{(i)}, \Delta \rangle \right) \exp \left( k \langle \phi_j^{(i)}, \Delta \rangle \right).
    \label{eq:3D_uitleg}
\end{equation}
In other words, on each data line the model consists of only $n_i$ terms, while in the convex hull of the parallel lines it consists of $n_0 + \ldots n_{l-1}$ terms. Remember that all of $l$ and $n_0, \ldots, n_{l-1}$ are small integer numbers. 

From Equation \eqref{eq:3D_uitleg} and Equation \eqref{eq:Sparse_interpolant} we consequently find
\begin{multline}
    b_j^{(i)}(k\Delta+\delta^{(i)}) = \beta_j^{(i)} \exp \left( -p(\delta^{(i)}) \langle \phi_j^{(i)}, \Delta \rangle \right), \\ k=0, \ldots, N_i-1, \quad i=0, \ldots, \ell-1, \quad j=1, \ldots, n_i.
    \label{eq:3D_model}
\end{multline}
Note that $b_j^{(i)}(U)$ remains constant along each line of the form $R+p(U)\Delta$ and only varies with the projection $R$ of that line onto the normal plane.

Remains to determine the $b_j^{(i)}(u, v, w)$. These functions can be determined from the interpolation conditions given in Equation \eqref{eq:3D_model}. A simple model for $b_j^{(i)}(u, v, w)$ is a 2D polynomial interpolant $a_j^{(i)}(p,q)$, as we outline now. Let us denote the coordinates of the intersection point given in Equation \eqref{eq:Intersection_point} by $R=(r, s, t)$. The collection of points on a particular line perpendicular to the normal plane, say here through $R$, is entirely identified by the remaining two degrees of freedom that pinpoint the intersection point of such a line with the normal plane. Since every  point $U=(u,v,w)$ on the line perpendicular to the normal plane and passing through $R$ satisfies the conditions
$$\frac{u-r}{\Delta_u} = \frac{v-s}{\Delta_v} = \frac{w-t}{\Delta_w},$$
we can take any two of the values
\begin{equation}
\begin{split}
    &\Delta_v u -\Delta_u v = r\Delta_v - s\Delta_u \\
    &\Delta_w u -\Delta_u w = r\Delta_w - t\Delta_u \\
    &\Delta_w v -\Delta_v w = s\Delta_w - t\Delta_v
\end{split}
\label{eq:line_characterization}
\end{equation}
to characterize the full line. Over the whole of such a perpendicular line
the right hand sides of Equation \eqref{eq:line_characterization} are constant and independent of the points $U$ on the line. The right hand sides of Equation \eqref{eq:line_characterization} are only determined by $\Delta$ and $R$. Say, for now, that we take the first two of \eqref{eq:line_characterization}, without any loss of generality: $p = \Delta_v u -\Delta_u v, q = \Delta_w u -\Delta_u w$. For the $l$ parallel lines on which the samples where collected, we find
$$\left( p^{(i)}, q^{(i)} \right) = (\Delta_v \delta_u^{(i)} -
\Delta_u \delta_v^{(i)}, \Delta_w \delta_u^{(i)} - \Delta_u
\delta_w^{(i)}), \qquad i=0, \ldots, l-1.$$

Let us abbreviate the values in the right hand side of Equation \eqref{eq:3D_model} by $\alpha_j^{(i)}$ and replace $b_j^{(i)} (u,v,w)$ in Equation \eqref{eq:3D_model_2} by the more appropriate
$a_j^{(i)}(p,q)$ since the interpolation conditions for $b_j^{(i)}(u,v,w)$ hold for a whole line and vary only with the intersection point of such a line with the normal plane:
\begin{equation}
    T_{RMS}(u,v,w) \approx \sum_{i=0}^{l-1} \sum_{j=1}^{n_i}
    a_j^{(i)}(p,q) \exp \left( p(u,v,w) \langle \phi_j^{(i)}, \Delta \rangle \right).
    \label{eq:blended model}
\end{equation}


Finally, the 2D polynomial interpolant  $$a_j^{(i)}(p,q) = \sum_{h, \ell} \tau_{h\ell}^{(i,j)} T_h(p) T_\ell(q)$$ 
where $T_n(\cdot)$ denotes the well-known Chebyshev polynomial  (of the first kind) of degree $n$,
is computed from the interpolation conditions
$$a_j^{(i)}\left( p^{(m)}, q^{(m)} \right) = \left\{ \aligned 
&\alpha_j^{(i)}, \qquad m=i \\ &0, \qquad m \not= i,  
\endaligned \right. \qquad i, m=0, \ldots, l-1, \qquad j=0, \ldots, n_i.$$

\begin{figure}[h]
    \centering
    \includegraphics[width=0.49\columnwidth]{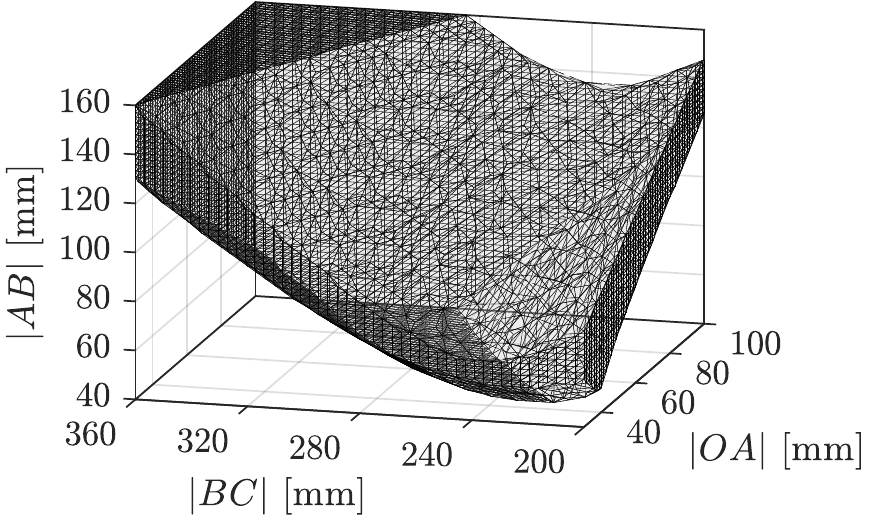}
    \includegraphics[width=0.49\columnwidth]{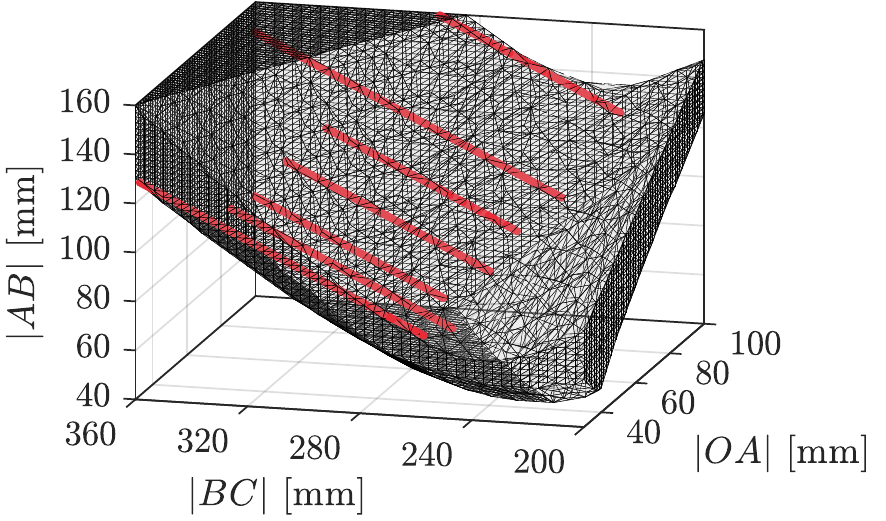}
    \caption{Region of interest delimited by \eqref{eq:Static_constraint_psi} and \eqref{eq:Dynamic_constraint_psi} at the left and sampling locations on $l=7$ parallel lines at the right in red.}
    \label{fig:sampling_method}
\end{figure}

\begin{figure}[h]
    \centering
    \includegraphics[width=0.48\columnwidth]{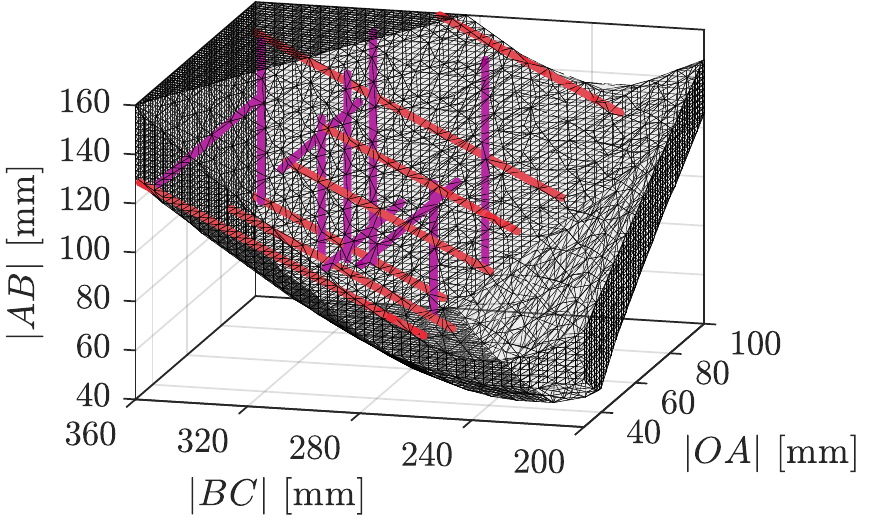}
    \includegraphics[width=0.48\columnwidth]{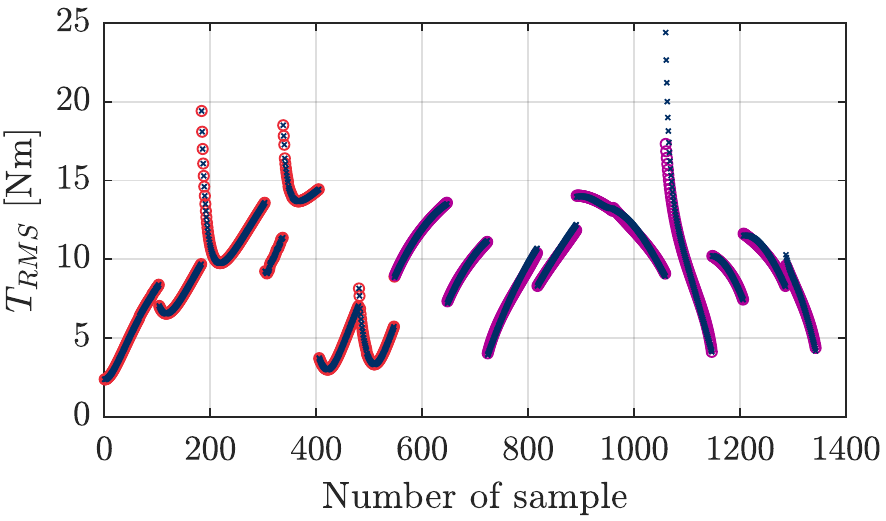}
    \caption{Validation directions at the left in purple and validation results of the blended model \eqref{eq:blended model} at the right.}
    \label{fig:validation}
\end{figure}

We now apply the above to our four-bar problem. The region of interest for the design variables $|OA|, |AB|, |BC|$ and restricted by the conditions \eqref{eq:Static_constraint_psi} and \eqref{eq:Dynamic_constraint_psi} is shown in Figure \ref{fig:sampling_method} and the sampling performed in this region is shown in red in Figure \ref{fig:sampling_method} (right). We take $l=7$ and $\Delta=(0.000, 0.920, 0.503)$ to guarantee maximal coverage of the region of interest. Also, the whole domain is translated over $-(31.000, 257.859, 72.705)$ to start sampling at the origin, in line with our description. In total only 618 samples are determined by the simulations explained in Section \ref{sec:CAD motion simulations}, which shape the objective function. We find that $n_i=5$ for all $i=0, \ldots, 6$, thus yielding $7 \times 5$ terms in the global model
$T_{RMS}(|OA|, |AB|, |BC|)$. The coefficients $a_j^{(i)}(p,q)$ are interpolated by a linear combination of the 7 bivariate Chebyshev polynomials $T_m(p)T_n(q), 0 \le m+n \le 2$ and $T_2(p)T_1(q)+T_1(p)T_2(q)$. As a final step, we validate the blended model by collecting 1252 more simulation data on 10 other lines within the convex hull, along directions different from $\Delta$. These evaluation directions are shown in purple in Figure \ref{fig:validation} (left) and the result of this validation is shown in Figure \ref{fig:validation} (right). In Figure \ref{fig:validation} (right) the red and purple markers depict the simulated data and the blue markers represent the value computed from the blended model \eqref{eq:blended model}. Each partial curve shows the function values of $T_{RMS}(u,v,w)$ restricted to one of the lines where samples were collected, either for interpolation (red) or validation (purple). The overall Root Mean Square Error (RMSE) equals 0.0281Nm, indicating a very good fit. When restricting our attention to $T_{RMS}$ values below 5 -- reasonable to locate a minimum -- the RMSE reduces to 0.0153Nm.

After this validation we look for a minimum of the modelled $T_{RMS}(u,v,w)$ \eqref{eq:blended model} in the convex hull of the parallel lines shown in Figure \ref{fig:sampling_method} (right). This was fulfilled through a brute force evaluation of the objective function \eqref{eq:blended model}, in which 10,000,000 calculations were performed in 3 minutes. Thus, we can conclude that the most time consuming was the collection of all 1870 samples, in which generating a sample takes on average 1 minute and 25 seconds of simulation time. The model reaches a minimal value of 2.5989Nm at $U=(33.246, 266.088, 79.435)$. If we want to achieve the same result through brute force evaluation of the simulations, 10,000,000 simulations would be required.

\section{Conclusion}

\label{Conclusion}

This study proposes an industrially applicable approach that guarantees to reveal the global optimal design of a four-bar mechanism based on CAD motion simulations and sparse interpolation. The process of sampling the objective-value $T_{RMS}$ for a combination of design parameters $\vert OA \vert$, $\vert AB \vert$ and $\vert BC \vert$ is automated by means of CAD multi-body motion simulations. Subsequently, the constraints limiting the feasible design space are introduced based on the position analysis of the four-bar mechanism. This guarantees that all designs considered by the optimiser can be assembled, and no circuit or branch defect will occur during the mechanism's movement.

If the unconstrained design space would be considered by a brute force approach 10,000,000 objective value samples would be required. As each objective value sample requires a simulation of approximately 1 minute and 25 seconds this would be practically impossible and seriously hampering the identification of the global optimum. However, thanks to the mathematical description of the design space constraints, introduced in this paper, sparse interpolation can be applied. The innovative sparse interpolation technique, described and applied in this paper reduces the number of necessary simulations to only 618. This allowed to identify the global optimal design. As shown in Table \ref{tab:Results_designoptimization}, the method clearly outperforms the best result (local optimum) obtained through the HEEDS Sherpa heuristic optimizer \cite{RedCedarTechnology2014}. The global optimum is 38 \% more efficient than the local optimum. Moreover, the global optimum also reduces the $T_{max}$ by 67 \% compared to the original design, which means that the mechanism can operate with a smaller, and thus cheaper motor.
\begin{table}
\centering
\caption{Saving potential achieved with design optimization.}
\label{tab:Results_designoptimization}
\begin{tabular}{cccccccc}
\textbf{Design}                                         & \begin{tabular}[c]{@{}c@{}}\textbf{$\vert OA \vert$}\\\textbf{[mm]}\end{tabular} & \begin{tabular}[c]{@{}c@{}}\textbf{$\vert AB \vert$}\\\textbf{[mm]}\end{tabular} & \begin{tabular}[c]{@{}c@{}}\textbf{$\vert BC \vert$}\\\textbf{[mm]}\end{tabular} & \begin{tabular}[c]{@{}c@{}}\textbf{\textbf{\textbf{\textbf{$T_{rms}$}}}}\\\textbf{\textbf{\textbf{\textbf{[Nm]}}}}\end{tabular} & \begin{tabular}[c]{@{}c@{}}\textbf{\textbf{$T_{max}$}}\\\textbf{\textbf{[Nm]}}\end{tabular} & \begin{tabular}[c]{@{}c@{}}\textbf{\textbf{\textbf{\textbf{$T_{rms}$}}\textbf{\textbf{}}}}\\\textbf{\textbf{\textbf{\textbf{savings}}}}\\\textbf{\textbf{\textbf{\textbf{[\%]}}}}\end{tabular} & \begin{tabular}[c]{@{}c@{}}\textbf{\textbf{\textbf{\textbf{\textbf{\textbf{\textbf{\textbf{$T_{max}$}}}}\textbf{\textbf{\textbf{\textbf{}}}}}}}}\\\textbf{\textbf{\textbf{\textbf{\textbf{\textbf{\textbf{\textbf{savings}}}}}}}}\\\textbf{\textbf{\textbf{\textbf{\textbf{\textbf{\textbf{\textbf{[\%]}}}}}}}}\end{tabular}  \\ 
\hline
Original                                          & 53                                                                               & 65                                                                               & 282                                                                              & 7.91                                                                                                                           & 13.26                                                                                      & -                                                                                                                                                                                              & -                                                                                                                                                                                                                                                                                                                             \\
\\
\begin{tabular}[c]{@{}c@{}}Local\\optimum\end{tabular}  & 40.6                                                                             & 77.2                                                                             & 263.23                                                                           & 4.19                                                                                                                           & 6.30                                                                                       & 47                                                                                                                                                                                             & 52.5                                                                                                                                                                                                                                                                   \\                                                         \\
\begin{tabular}[c]{@{}c@{}}Global\\optimum\end{tabular} & 33.246                                                                           & 79.435                                                                           & 266.088                                                                          & 2.60                                                                                                                           & 4.35                                                                                       & 67                                                                                                                                                                                             & 67                                                                                                                                                                                                                                                                                                                           
\end{tabular}
\end{table}

\newpage

 \bibliographystyle{elsarticle-num} 
 \bibliography{library}





\end{document}